# A New Determination Of The Diffuse Galactic and Extragalactic Gamma-Ray Emission

Andrew W. Strong[1], Igor V. Moskalenko[2,3] & Olaf Reimer[4]

[1]*Max-Planck-Institut für extraterrestrische Physik, 85740 Garching, Germany*
[2]*NASA Goddard Space Flight Center, Code 661, Greenbelt MD 20771, USA*
[3]*Joint Center for Astrophysics, University of Maryland, Baltimore MD 21250, USA*
[4]*Ruhr-Universität Bochum, TP IV, 44780 Bochum, Germany*

**Abstract.** The GALPROP model for cosmic-ray propagation is able to make explicit predictions for the distribution of galactic diffuse gamma-rays. We compare different propagation models with gamma-ray spectra measured by EGRET for various regions of the sky. This allows sensitive tests of alternative explanations for the apparent excess emission observed at GeV gamma-rays. We find that a population of hard-spectrum gamma-ray sources cannot be solely responsible for the excess since it also appears at high latitudes; on the other hand a hard cosmic-ray electron spectrum cannot explain the gamma-ray excess in the inner Galaxy. By normalizing the cosmic ray spectra within reasonable bounds under preservation of their shape we are able to obtain our best prediction of the Galactic component of diffuse gamma rays, and show that away from the Galactic plane it gives an accurate prediction of the observed gamma-ray intensities. On this basis we reevaluate the extragalactic gamma-ray background. We find that for some energies previous work underestimated the Galactic contribution and hence overestimated the background. The new EGRB spectrum shows a positive curvature similar to that expected for models of the extragalactic gamma-ray emission based on contributions from unresolved blazars.

## INTRODUCTION

Diffuse continuum gamma-rays from the interstellar medium are potentially able to reveal much about the sources and propagation of cosmic rays, but in practice the exploitation of this connection is problematic. We have previously [5] compared a range of models, based on our cosmic-ray propagation code GALPROP, with data from experiments flown aboard the Compton Gamma Ray Observatory. While it is rather easy to get agreement within a factor ~2 from a few MeV to a few GeV with a “conventional” set of parameters, the data quality warrants considerably better fits. Specifically, in order to reproduce the apparent excess emission at GeV energies observed by EGRET, we adopted a model with a broken electron injection spectrum and a proton spectrum preserving its shape but deviating moderately from that measured locally. With this model we were able to reproduce the gamma-ray spectrum at all sky regions as well as the longitude and latitude projections of the EGRET measured gamma-ray data as a function of energy. An essential feature of our approach was that the locally-measured cosmic-ray spectrum of electrons is not a prominent constraint because of the spatial fluctuations due to energy losses and the stochastic nature of the sources in space and time; the average interstellar electron spectrum responsible for gamma-rays via Inverse Compton emission can therefore be quite different from that measured locally. A hard electron spectrum interpretation has been found not convincing, since the electron fluctuations required are larger than expectable. In addition, more accurate measurements of the local cosmic-ray particle spectrum allow less freedom for deviations in the $\pi^0$-decay component. If large variations in the proton spectrum are postulated as explanation for the observed GeV excess emission, we showed that this leads to severe overproduction of secondary antiprotons and positrons, so that a hard nucleon spectrum possibility is effectively excluded. Another suggestion which has been made is that the gamma-ray spectrum contains a $\pi^0$-decay component from cosmic-ray protons close to their (SNR) sources [1], so that in gamma-rays we see an injection spectrum which is much flatter than the propagated spectrum which we measure locally. A shortcoming of previous analyses was that the comparison with EGRET data was limited to particular sky regions only so the rich EGRET data have remained not fully exploited. Here we attempt to investigate the fact that the models predict quite specific behavior for different sky regions, and this provides a sensitive test: the best model should be consistent with the data in all directions (sky regions) where different propagation models have been compared with the data. We demonstrate that an optimized model, with less dramatic changes in the electron and nucleon spectra relative to a conventional model, can well reproduce the gamma-ray data, over the entire sky and up to 100 GeV in energy.

## DATA AND METHOD

We use EGRET counts and exposure all-sky maps in Galactic coordinates with 0.5° binsize; gamma-ray point sources of the 3EG catalogue have been removed; finally, the energetic range has been expanded up to 100 GeV. For the spectra, the statistical errors on the EGRET data points are very small since the regions chosen have large solid angle; the systematic error dominates and we have conservatively adopted a uncertainty of 15% in plotting the observed spectra. The predicted model sky maps are convolved with the energy dependent EGRET point-spread function. The spectra are compared in the sky regions summarized in Table 1.

**TABLE 1. Sky regions as chosen for evaluation with GALPROP and EGRET data**

| **Region** | **l** | **\| b \|** |
|---|---|---|
| A: Inner Galaxy | 330° - 30° | 0 – 5° |
| B: Galactic plane avoiding inner Galaxy | 30° - 330° | 0 – 5° |
| C: Outer Galaxy | 90° - 270° | 0 – 5° |
| D: Intermediate latitudes region 1 | 0 – 360° | 10° - 20° |
| E: Intermediate latitudes region 2 | 0 – 360° | 20° - 60° |
| F: Galactic poles | 0 – 360° | 60° - 90° |
| G: D+E+F for EGRB determination | 0 – 360° | 10° - 90° |
| H: Region as used in [2] | 300° - 60° | 0 – 10° |

All models use the locally-observed proton and Helium spectra. This is done because the nucleon data are now more precise than those which were available to us [4] in 2000. The nucleon injection spectra and the propagation parameters are chosen to reproduce the most recent measurements of primary and secondary nuclei. The radial distribution of cosmic-ray sources used is the same as in [5], since we find this empirically-derived form still gives a good reproduction of the spatial gamma-ray distributions. Although flatter than the SNR distribution, this may be compensated by the gradient in the CO-to-$H_2$ conversion factor which has a metallicity and temperature dependence, such having the net effect of causing the factor to increase with radius. A more detailed examination of this effect is given in [8]. Here a uniform value of $X_{CO}=1.9 \times 10^{20}$ molecules $cm^{-2}$/ (K km $s^{-1}$) has been used. The halo height is taken as 4 kpc as in [5].

## MODELS

Initially, we repeat the test of a "conventional" model which is based on the locally-observed electron (as well as nucleon) spectrum. The model spectra are compared with EGRET data in the sky sections as given in Table 1, for various energy regimes and in latitudinal and longitudinal projections. As reported in previous work, such a model shows an excess the GeV region relative to that predicted; what is now evident is that this excess appears in *all* latitudes/longitude ranges. This already shows that the GeV excess is not a feature restricted to the Galactic ridge or the gas-related emission. Further it is clear that a simple upward rescaling of the $\pi^0$-decay component will not improve the fit in any region, since the observed peak is at higher energies than the $\pi^0$-decay peak. Since the spectrum is very different from that of a $\pi^0$-decay even at intermediate latitudes, a substantial Inverse Compton component is required. In the "SNR source" scenario [1] the spectrum in the inner Galaxy is attributed to an additional population of unresolved SNR, but this component cannot explain the excess at high latitudes, and hardly in the outer Galaxy. This explanation is therefore by itself insufficient, although it could be part of the solution.

Our approach is to concentrate on obtaining a fit in all regions apart from the inner Galaxy, since then we can be reasonably sure that no dominant population of unresolved sources is distorting the spectrum, and individual sources would be nearby, rare and easily identified, and so the situation should be relatively simple. A hard electron spectrum interpretation is effectively excluded, since the required fluctuations in the electron spectrum exceed what can be considered as still acceptable and contradicts the spectral data above 10 GeV. Since the Galactic intensities are lower away from the inner Galaxy, the extragalactic component is critical. Although the Inverse Compton contribution (via the electron spectrum) is the main unknown, and the EGRB is always adjustable to fit at the poles (where the total predicted Galactic + EGRB has to agree with the data), we adopt an optimization procedure aimed to produce a continuous spectrum for combining with the model Galactic components. Hence, the analysis is not sensitive to details of the EGRB.

An example of the result of our optimization procedure is shown in Fig.1 and 2. An particle injection spectrum with power-law index of 1.5 (< 20 GV for electrons; < 10 GV for protons), and 2.4 (> 20 GV for electrons; > 10 GV for protons) is found optimal, consistent with what we reported earlier [5]. Values differing by up to 0.1 from this would also be acceptable. Overall the agreement is satisfactory and much improved over the conventional model. The spectra in different sky regions are satisfactorily reproduced and there is no longer an GeV excess apparent. Hence the diffuse emission spectrum can now be reproduced from 30 MeV to 100 GeV. The proposed scenario implies a substantial contribution from Inverse Compton at all energies, making Inverse Compton the dominant emission component at |b| > 10° at all energies. Also, the optimized model reproduces the cosmic ray antiproton and positron spectra well [6].

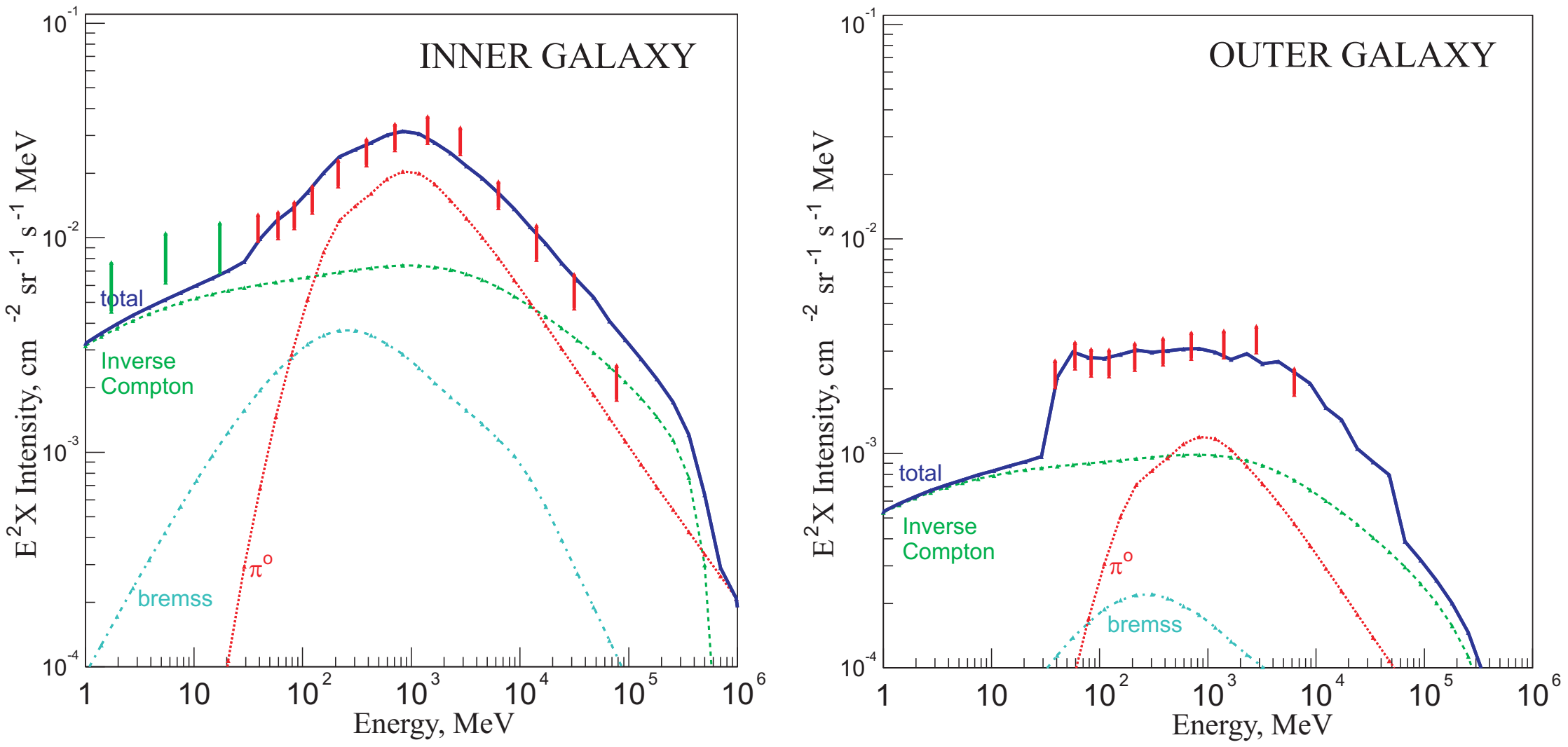

**FIGURE 1.** Spectra of the diffuse galactic $\gamma$-ray emission in the inner and outer galaxy in the optimized model. Introducing a relatively mild deviation from the locally measured electron and proton spectra a good agreement with the EGRET data from 30 MeV to 100 GeV has been achieved.

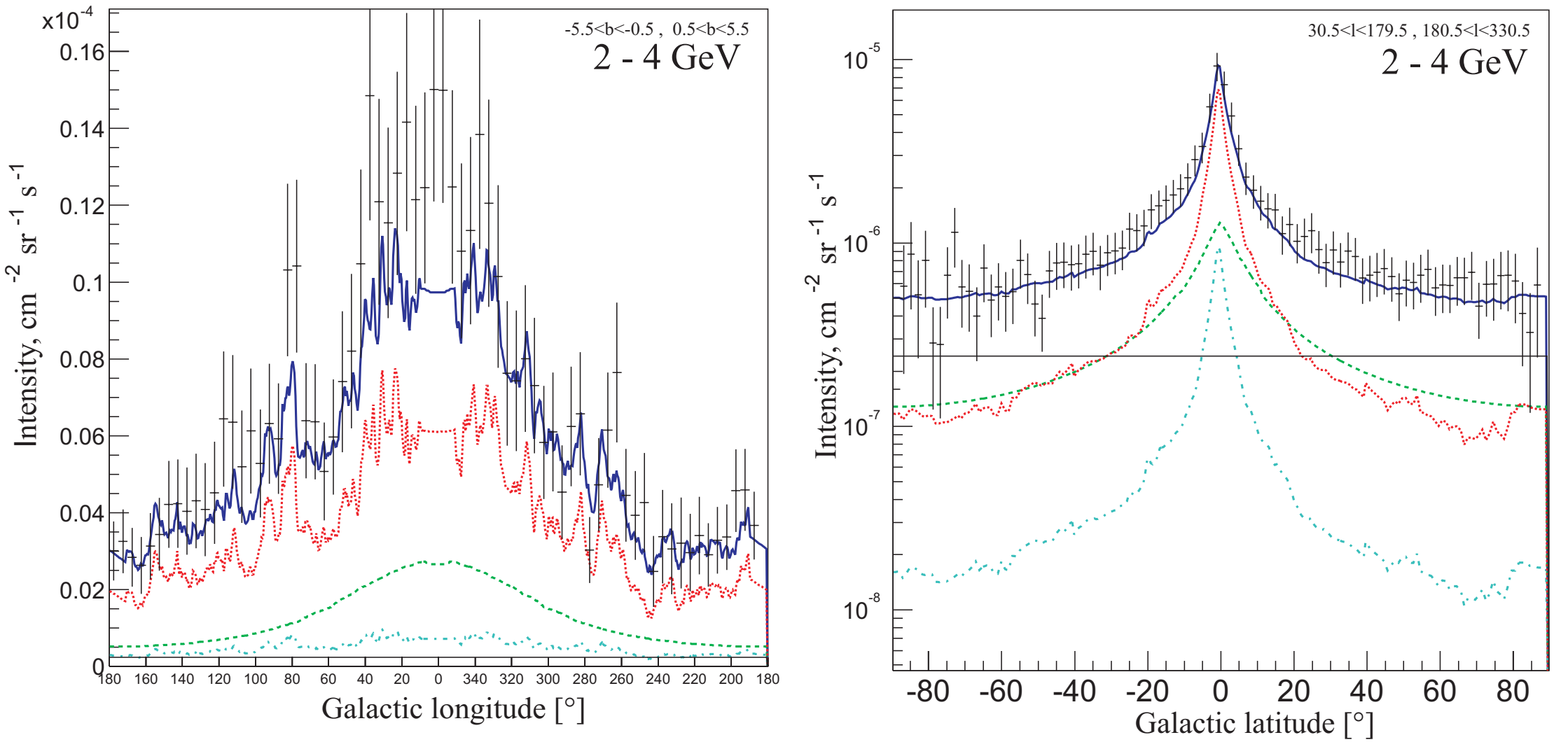

**FIGURE 2.** Longitude and latitude profiles as predicted in the optimized model, compared with EGRET data in the energy range from 2 to 4 GeV where the "GeV-excess" has been most prominently seen previously. Coding of the individual components as in Fig.1.

The longitude profiles are also mostly satisfactory, considering that the model does not aim to include details of Galactic structure yet (e.g., spiral arms), and the systematic deviations reflect the lack of an exact fit to the spectra. Latitude profiles show excellent agreement with the EGRET data in all energy ranges, which in our opinion gives the most convincing support for this interpretation. In particular, the reproduction of the high-latitude variation confirms the existence of a dominant Inverse Compton component.

## EXTRAGALACTIC DIFFUSE GAMMA-RAY BACKGROUND

Given the success of the model [5] in reproducing the gamma-ray sky, we can use it to determine improved estimates of the EGRB. Since the model is nevertheless not exact the best approach is to fit the observed intensities with a free scaling factor; in this way the EGRB is determined as the intercept, thus removing any residual uncertainty in the absolute level of the Galactic components. This is essentially the same method as used in [4], the difference lies in the diffuse emission model. To reduce the effects of Galactic structure, unresolved point sources etc. the fits are made excluding the plane, using the region G; ideally both IC and gas-related components would be left free but they are difficult to separate statistically at high latitudes, so we make a linear fit to the total IC+$\pi^0$-decay+bremsstrahlung, with the scaling factor and EGRB as parameters. The fit and uncertainties are based on a simple $\chi^2$ analysis, with (l,b) bins 360° x 2° to obtain sufficient statistics (at least 10 counts per bin were required). For comparison we also made fits to the entire sky; in this case IC and gas-related contributions are easily separated; the two fit regions then give some indication of the model-dependent systematic error in our EGRB estimates. The scaling factors determined for the region G fits reflect the deviations from the model and are typically between 0.8 and 1.2, which is satisfactory. For 30-50 MeV, 1-2 and 2-4 GeV the scaling factors deviate further from unity reflecting the discrepancy in the spectrum so that these are the least reliable ranges of our EGRB determination. The EGRB is however not very sensitive to the scaling factor.

The two fitted regions (region G and all-sky) give consistent results, indicating that there is no large systematic effect; it shows a model-dependent systematic uncertainty of 5-30%, comparable to the formal statistical errors. This is comparable to the ~15% systematic uncertainty on EGRET data so we adopt 30% for our total error estimate. Fig. 3 shows the extragalactic X- and $\gamma$-ray background, using the compilation by [4] but with our new determination from the EGRET data, and also updated COMPTEL results [9]. Our estimates lie significantly below those of [4], in most energy ranges. A positive curvature in our EGRB spectrum is indicated which is not unexpected [2] but in view of the systematic uncertainties should not be taken too literally; a similar, less pronounced effect is already present in previously reported EGRB spectrum [4]. Although the 50 MeV - 2 GeV range can be fit satisfactorily by a power law, this is clearly inconsistent with the data extending beyond 2 GeV. The reason for the difference between our spectrum and that given previously on the basis of the EGRET data lies primarily in the improved modeling of high-latitude gamma-rays based on Inverse Compton emission from the halo.

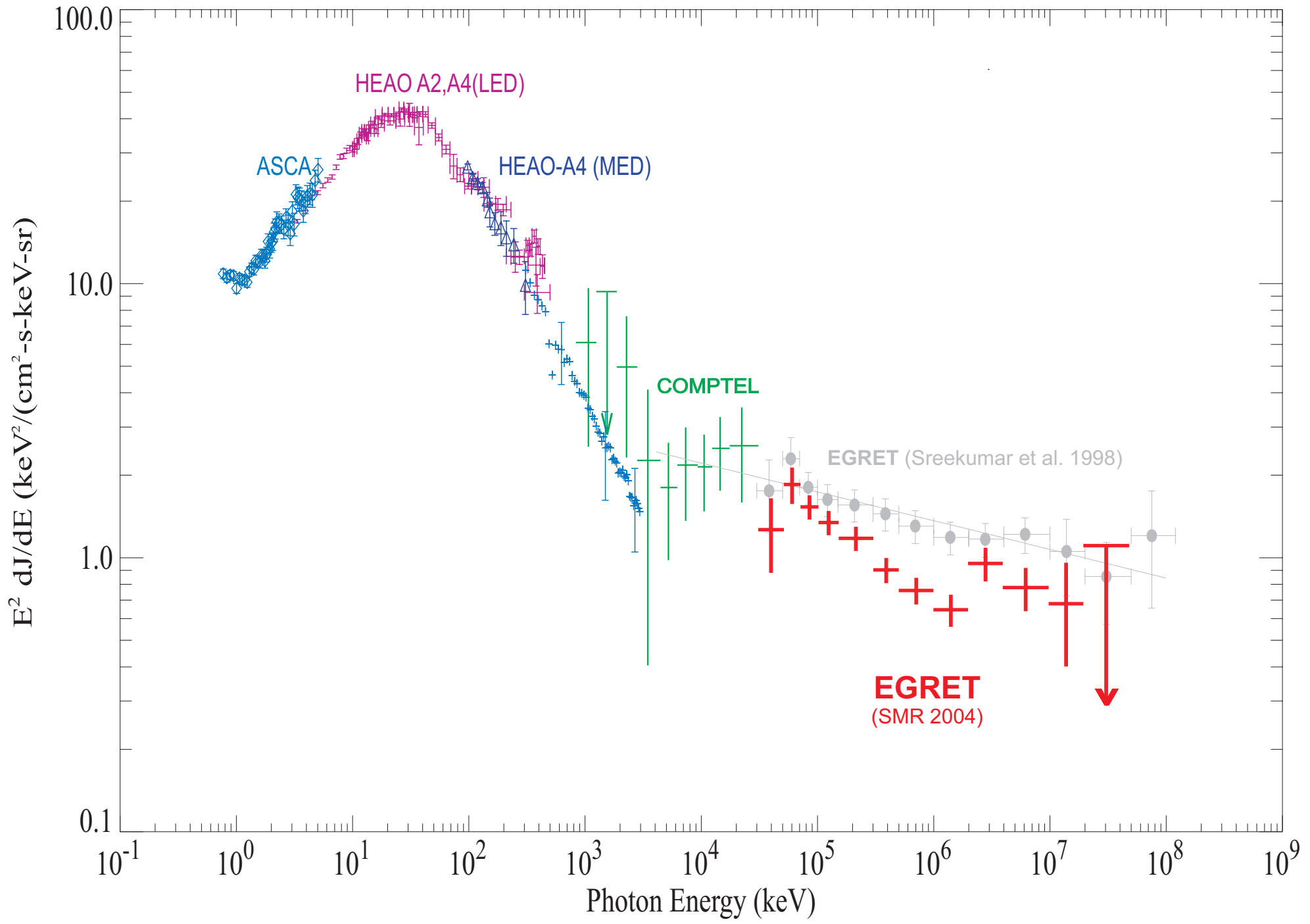

**FIGURE 3.** The spectrum of the extragalactic X-ray and γ-ray background, contrasting the previously reported spectrum [4] with that from our recent re-evaluation [7]. Data compilation from [3] except COMPTEL from [8].

# ACKNOWLEDGMENTS

We would like to particularly thank David Bertsch for the assistance and discussions on the subject of the events and instrumental response of the EGRET telescope above 10 GeV and Seth Digel for providing the kinematically analyzed HI and CO data. I.M. acknowledges partial support from a NASA Astrophysics Theory Program grant. O.R. acknowledges support from the BMBF trough DLR grant QV0002.